# Glassy nature of hierarchical organizations

Maryam Zamani[b] and Tamas Vicsek[a,b]

[a]Department of Biological Physics, Eötvös University, Pázmány P. stny 1/A, 1117 Budapest, Hungary, and [b]Statistical and Biological Physics Research Group of the Hungarian Academy of Sciences, Pázmány P. stny 1/A, 1117 Budapest, Hungary

*Abstract*

The question of why and how animal and human groups form temporarily stable hierarchical organizations has long been a great challenge from the point of quantitative interpretations. The prevailing observation/consensus is that a hierarchical social or technological structure is optimal considering a variety of aspects. Here we introduce a simple quantitative interpretation of this situation using an approach reminiscent of those developed for describing complex behaviour in terms of statistical mechanics. We look for the optimum of the efficiency function $E_{eff} = 1/N \sum_{ij} J_{ij} a_i a_j$ with $J_{ij}$ denoting the nature of the interaction between the units $i$ and $j$ and $a_i$ standing for the ability of member $i$ to contribute to the efficiency of the system. Notably, this expression for $E_{eff}$ has a similar structure to that of the energy as defined for spin-glasses. There is, however, an essential and novel feature of our approach: instead of optimizing by looking for a locally optimal state of the units in the nodes of a pre-defined network, we search for extrema in the complex efficiency landscape by finding locally optimal network topologies using a standard Monte Carlo method.

-------------------------------------------------------------

## I. Introduction

Hierarchy is obviously one of the most widespread features of natural, technological and social systems[1-4]. The behaviour of these systems is typically complex and their most relevant organizational principle is that the ties among the units they have correspond to an underlying network displaying hierarchical features. One of the important related questions is concerned with the driving forces that make the network topology converge to a single hierarchy (representing one of the many possible locally optimal structures). This paper is about developing and studying a simple approach which is capable of reproducing the emergence of a multi-level network structure based on the degree to which the units (individuals) are able to contribute to the efficiency (capacity to operate on a high level) of the system. We shall adopt terminologies on one hand used in statistical mechanics and network science, while, on the on the other hand, being typically used in the context of organizations and the underlying networks of collaborations. However, we expect this framework to be applicable to a significantly larger class of systems. Thus we consider the groups of humans as a paradigm, but our approach is so general that it is expected to be applicable to simpler systems such as groups of collaborating animals (apes, wolfs, etc.) as well as complex machines constructed by people.

The main novelty of our concept, which we point out here, before going into its details, that our structures have nodes with pre-defined abilities, are connected with directed edges having also a sign associated with them (positive and negative signs corresponding to collaboration or antagonistic relations, respectively). Searching for optimal states then is carried out by modifying the network topology so that both the collaborating partners and the flow of influences result in a maximal efficiency. This is contrasted with the standard statistical mechanics approach to complex systems made of a fixed set of interactions and a variable state of the nodes (i.e., as it is assumed in the case of the so called spin-glasses (see Refs. [5-7] for reviews). Having a relatively simple model of the emergence of hierarchical structure allows the study of several essential questions, for example the stability and the adaptability of realistic networks.

Below we develop an approach to address the question of the spontaneous emergence of hierarchical networks displaying behaviours analogous to those of glasses. By glassy behaviour we mean that while we are searching for a stable state, our structures do not converge to a unique network with a well-defined extremal value of their efficiency (an analogue of the energy in the physics literature: instead, the system "freezes" into various disordered structures representing a local extremum and full of strains or frustrations). Our main assumption is that such a structure should be optimal from the point of the efficiency of an organization (otherwise more efficient structures would take over in the situation in which organizations - even such as universities - compete). Prior works have shown that hierarchical organization can be advantageous, but either have not addressed the network aspect of the dominance[8] hierarchies or considered the embedded[9-11] cases. A few very recent studies considered more complex situations and proposed relatively complicated approaches to the treatment of the emergence of the kind of multi-level hierarchical networks containing directed edges, i.e., flow-hierarchies, in which individuals in various societies (including both human and animal) are typically situated[12,13]. A number of further interesting and remarkable aspects (e.g., origins) of hierarchies have recently been explored in the quickly growing literature on the topic[14-20]

## II. Modelling organizations

Since hierarchy is so abundant, we take an approach that assumes as little details as possible about the interactions of the units and about the function $E_{eff}$ that is associated with the efficiency (or fitness, performance, etc.) of the group of entities forming the group/organization. This has to be so, because many and very different "rules" are unlikely to lead to the same universal behaviour. At first, we define the basic quantities we shall use when specifying the process of obtaining an efficient organization.

We represent the relations in an organization by a network made of *directed edges* standing for the leader-follower relations in the system (this can be generalized to undirected edges, but here we confine our study to the directed case). A group of $N$ members having $M$ connections among them (if we do not exclude the set of not fully connected subgraphs), has a total number of possible states equal to

$$\text{Number of configurations} = \frac{(N(N-1)/2)!}{(N(N-1)/2-M)!M!}2^M \quad , \tag{1}$$

where $2^M$ is the number of choices of the set of directions one can have for one of the $(N(N-1)/2)!/(N(N-1)/2-M)!M!$ possible configurations of (yet undirected) edges. In an ideal case  the direction of an edge between members $i$ and $j$ would point from $i$ to $j$ if $a_i > a_j$ (it is advantageous and is typically indeed the case that agents with higher abilities can enforce their decisions on agents with smaller abilities, i.e., occupy a higher position within the organizational hierarchy. However, with some finite probability, in a realistic case a proportion $p$ of all of the links between two members points from the less knowledgeable to the more knowledgeable person (for several reasons, such as personality, i.e., a "boss" has smaller information than a subordinate, etc.)

Next we consider the contribution of a pair of interacting agents to the successful operation of the organization. We make the plausible assumption that on an absolute value scale their contribution is between 0 and l. In addition, $E_{eff}$ is linearly proportional to their abilities, i.e., $E_{eff} = a_i\ a_j$. However - and this is an essential point, when one considers the relations of sophisticated creatures - the interaction between two individuals can be both harmonic with and antagonistic with a probability (1-$q$) and $q$, *respectively* (again, for several reasons, such as a prior or a persistent conflict, mismatch of personalities, etc.). In the "harmonic"case the contribution of the two members is positive, on the other hand, if they are in an antagonistic relation their interaction will result in a decrease of the total efficiency, thus their interaction enters the expression for the efficiency as negative contribution.

Assuming that the total performance of the organization can be represented as the contribution of the pairwise interactions we arrive at the (central for our study) expression

$$E_{eff}(p,q) = 1/N \sum_{ij}^{M} J_{ij}(p,q)a_i a_j \,, \qquad (2)$$

with the summation running over nodes that have at least one incoming or outgoing edge. According to the above arguments about the possible relations between two members, we assume that $J_{ij}$ can be equal to 1 or -1 (fruitful collaboration or harmful/antagonistic collaboration, respectively). This simple expression is rather similar to those describing spin systems (with $E$ being the energy, and $J$ being the strength of interaction between $i$ and $j$. There are two essential points that have to be stressed about eq. (2). However, when considering optimal performance of an organization we shall assume that the attributes (abilities) of the individual units are constant during the optimization. For the $a_i$ values we used randomly generated numbers on the unit interval following a bounded *log-normal distribution* (which can be argued to be characteristic for the outputs of complex entities (https://en.wikipedia.org/wiki/Log-normal_distribution)).

In this approach, the direction of the edges determines the structure or the "flows" of/on the directed graph, while the $J_{ij}$ values determine the total efficiency for a given configuration of the edges. Thus, an edge $ij$ has both a direction and a sign (a value 1 or -1) and the two are coupled in a probabilistic manner. Furthermore (see later), we shall have a constraint related to the number of edges a node can have.

(I) Although the expression (2) has a simple structure, we know from the statistical mechanics of spin-glasses that such expressions can result in extremely complex behaviour if $J_{ij}$ can have both positive and negative (antagonistic) values.

(II) Since in our scheme $a_i$ and $a_j$ are fixed (they represent attributes of individuals), *we are not optimizing (2) by suitably choosing the values of $a_i$: on the contrary, we optimize by finding the subset of M edges* (from the many possible sets) that, together with the corresponding $J_{ij}$ values result in the largest value of $E_{eff}$. In other words, we look for the optimal network (set of connected $i$ and $j$ members which maximizes the performance for the given $M$).

Thus, the only remaining task is to define $J_{ij}$ taking into account the complex nature of human behaviour in a simple manner.

The sign of $J_{ij}$ and the direction of the edge $ij$ are decided by two factors: 1) whether the $ij$ edge points from the larger to the lower ability of the participants $i$ and $j$ and 2) whether these participants are compatible or antagonistic. Thus, we choose (with the corresponding probabilities!)

i) $J_{ij}=1$ if the $ij$ edge points from a node with larger ability to a smaller $(a_i > a_j$ and the two individuals cooperate (and $J_{ij}= -1$ otherwise)

ii) $J_{ij}= -1$ if the $ij$ edge points from a node with smaller ability towards a larger one and the two individuals are antagonistic (and $J_{ij}=1$ otherwise)

If there is no edge between $i$ and $j$ then $J_{ij}=0$. As explained in the paragraph after eq. (1) the probabilities for $J_{ij}=1$ are $(1-p)(1-q)$ and $p(1-q)$ while for $J_{ij}= -1$ they are $(1-p)q$ or $pq$. $p$ and $q$ correspond to the probabilities of an inverse direction of the edge $ij$ (from $i$ to $j$ for $a_i > a_j$ and for an antagonistic collaboration, respectively. It is very important to stress at this point that we enforce the above rules hold for the subgaphs of $M$ edges as well! Of course, for simplicity, one can assume that $p=q$ and this would result in a single free parameter only.

iii) We consider a further essential restriction to make our system more realistic. In addition to the above, we also require that the total number of edges (incoming plus outgoing) from a node cannot exceed a pre-defined value $K$. We have to apply these constraints, otherwise the optimization would lead to unrealistic configurations consisting of cooperating members only, sub-graphs containing "top to down" edges and a nodes with a number of anomalously large number of edges (a person can manage only an approximately 4-10 incoming and outgoing relations since maintaining these is costly.) In fact, a simple calculation shows that without the above constraints the single maximum value of our $E_{eff}$ is equal to 1 for the case that abilities of individuals follow constant distribution. Instead of fixing $K$ beforehand, we first generate a subgraph of $M=3N$ edges and during an optimization we use as an upper limit for $K$ the number of edges that the node with the largest $K$ has.

## III. Simulations

We start out with a full graph with $N$ nodes each associated with a constant ability $a_i$ and with edges pointing towards lower ability sites from larger ability ones with a probability $1-p$. In addition, 1 or $-1$ is associated with each edge, independent of their direction (however, because of the term $1-p$, the number of negative edges for small $p$ will occur in larger overall number for the $a_i > a_j$ cases (than for $a_i < a_j$) so that the efficiency values and the structure of the graph become coupled. This full graph is time-independent; it contains all the information about the participants:

- their abilities $a_i$
- their willingness to collaborate pairwise with each other with $J_{ij}=1$ meaning they prefer to collaborate and $J_{ij}=-1$ if the two are not willing to collaborate.

We are searching for an optimal network, a subgraph containing $M$ edges within this graph of $N$ nodes (such that the number of nodes in the optimal network should be $N$). We start out with a random connected subgraph of $3N$ edges chosen from the many possible within the full graph of $N$ nodes. Throughout our calculations, the number of edges within the subgraphs we generate will satisfy the criteria that *in average* the ratio of edges in them pointing from a site with larger to a smaller ability will be equal to $1$-$p$ and the number of antagonistic interactions $J_{ij} = -1$ will be, again, in average, $qM$. This procedure introduces some noise into our data (as compared to considering configurations strictly satisfying the $1$-$p$ and the ratio of antagonistic edges equal to $q$ rules but this (otherwise "realistic") noise diminishes as $N$ is increased.

To satisfy iii) we allow new configurations only with nodes that have a total number links being smaller then $K$.

According to equation (1) the number of possible subgraphs increases with $M$ exponentially and - lacking an analytic solution, we explore the "efficiency landscape" numerically. The main quantity we shall be interested in is the probability density distribution of locally optimal efficiencies. This is a pragmatic approach, because knowing this distribution we can get an estimate how efficient is a given - existing in an organization - configuration relative to the achievable optimal ones. Of course, the probability that a simulation results in the absolute optimum is negligible due to the complex nature of our problem (because of the exponentially diverging possible states the number of optima also diverges).

When searching for networks with high efficiency, we start from a random configuration of typically $M = 3N$ edges embedded into our initial full graph and converge to a local maximum. According to the analogy with spin glasses, such local maxima have to exist and other maxima cannot be reached without rearranging the edges considerably. The procedure we follow is much like a Monte Carlo simulation, where efficiency plays the role of -energy. In each step a randomly selected edge is eliminated and next two random nodes are chosen which are not yet connected by one of the $M$-1 edges. The sign and the direction of the new edge, $J_{ij}$ is chosen according to conditions i)-iv) outlined in the *Modelling* section. First, we check whether the conditions A and B are satisfied, and if yes, the newly selected edge is added to the subgraph, while the one it replaces is eliminated. Then, using equation (2) the efficiency difference $\Delta E_{eff}$ between that belonging to the prior and the new configurations is calculated. The new network is accepted if $\Delta E_{eff} > 0$ and is also accepted if $\Delta E_{eff} < 0$ with a probability equal to exp (-$\beta$ $\Delta E_{eff}$ )), where $\beta$ is associated with external perturbations or noise. Sine we are interested in optimal, metastable states, we chose a relatively large value for $\beta$ being equal to 1000. If none of the two conditions related to $\Delta E_{eff}$ are satisfied, we drop this trial and choose a new alternative edge during the process. Fig. 1 shows the dependence of $E_{eff}$ as a function of the number of steps for three different initial conditions. It is clear from this plot that $E_{eff}$ saturates at varying values and after quite different number of steps. We assume that a local maximum has been achieved after $\Delta E_{eff} = 0$ for more than 10000 additional steps.

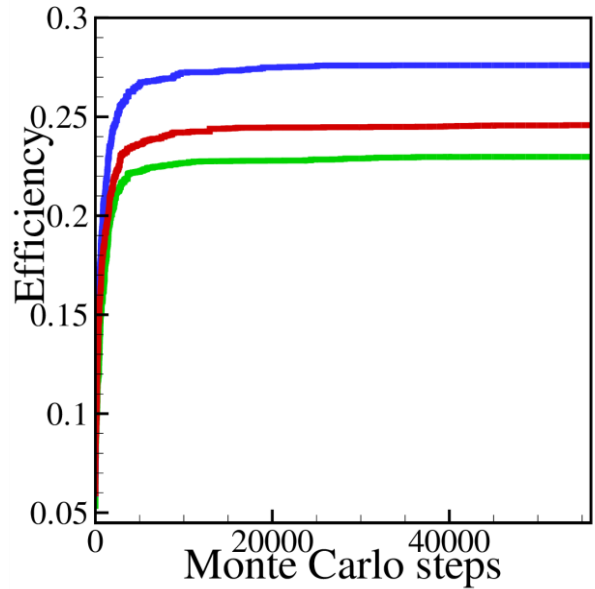

Figure 1. **Convergence of the calculated efficiencies to their locally stable (globally metastable) values.** Number of nodes $N$=64, $3N$ edges and for three different random initial conditions and for $p$=$q$=0.2 and $\beta$=1000.

## IV. Results

We concentrated our efforts to obtain results demonstrating our main point, i.e., that the cooperation of individuals emerging as a result of directed relations among them leads to non-trivial, locally optimal states. In other words, fully efficient structure cannot be achieved within reasonable computational time, and the final optimum - which can be obtained by a Monte-Carlo optimization - is dependent on both the initial conditions as well on the actual path of the trial and error process as we are attempting to approach a locally optimal state. We associate with a locally optimal state its efficiency and its network structure being characterised by the hierarchical nature of this network.

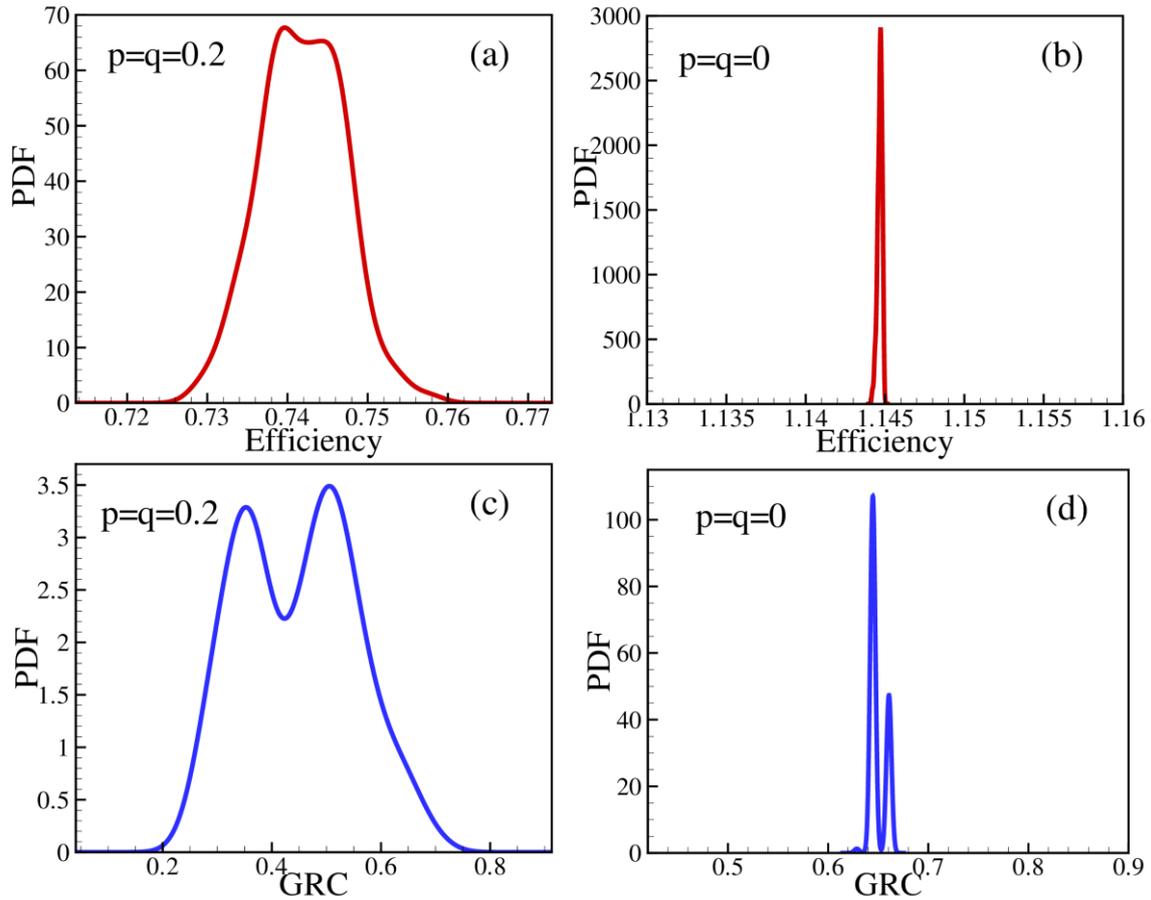

Figure 2. **Probability density function of efficiency and GRC (see text) values for one initial state.** $N$=64 and 3$N$ edges (a and c) $p$=$q$=0.2 (b and d) $p$=$q$=0, and no limitation for in and out-degrees, i.e., there is no frustration in the graphs for which the plots on the right were obtained.

Thus, at first we calculated the distribution of the efficiencies corresponding to the local maxima we obtained. The corresponding histogram (probability density function - PDF) was constructed for various $N$ values to see the size effect. The number of initial full graphs was 300 and 250 initial subgraphs of $M = 3N$ edges were used. Then, all of the obtained local optima were binned and a function was fitted to the resulting histogram. Fig. 2 demonstrates, that there is indeed a spread in the efficiency values of the available (locally) optimal networks. The system sizes we can consider are relatively small (as compared to some huge networks, e.g., on the internet), and there are two reasons for this. First, although we have carried out the calculations on a cluster of mini-supercomputers, the number of steps needed to obtain a figure is mostly huge (order of 300 x 250 x 40000). Second, we think our approach is more appropriate for organizations which have a number of individuals who are likely to be able to know each other, so have a number of members not larger than the Dunbar number (150, see, e.g., Ref. 21).

Thus, the systems we made the calculations for are of relatively limited size. However, the computational limitations we had to consider (the runs were carried out on a cluster of mini-supercomputers) were because we intended to explore the whole landscape of optima in the system. In a real, actual situation, there is – in principle – a simple search process only and the system settles down into the given optimum it ran into. Single runs – naturally – could have been executed for huge organizations as well.

Next we investigate the properties of the networks corresponding to the locally optimal states. Here we find a surprising result: the hierarchical nature of these networks remains rich and does not seem to converge to a single, simple configuration, close to its average.

We used to measure the level of hierarchy of the obtained networks by calculating the quantity called Global Reaching Centrality[22] (*GRC*). *GRC* is related to the distribution of the local reaching centralities, where the reaching centrality of a node is proportional to the number of nodes $C_R(i)$ which can be reached from node i one through the directed edges of the network. A spread (e.g., standard deviation) is a good indicator of the varying positions of the nodes within a hierarchical structure. (Many nodes can be accessed from the "top" ones and only a few from those at the "bottom"). Since the distributions of $C_R(i)$ is typically very different from a Gaussian, we chose to measure the spread of its values using the expression

$$GRC = \frac{\sum_{i \in V}[C_R^{\max} - C_R(i)]}{N-1} \qquad (3)$$

where $C_R^{max}$ is the largest of $C_R(i)$ and the summation goes over the nodes belonging to the graph V. GRC scores are strictly non-negative and the maximal *GRC* value of 1 is attained for star graphs. Regular and irregular trees also have a high *GRC* score[22], while Erdős-Rényi random networks attain a low *GRC* score of around $0.058 \pm 0.005$.

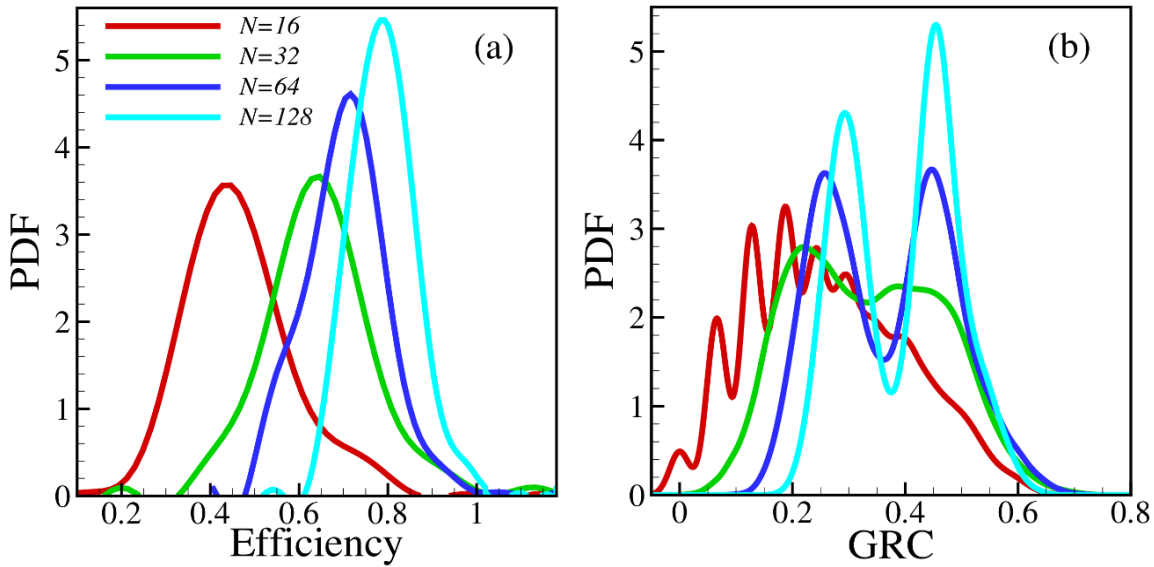

Figure 3. **Distribution of the local maxima of the efficiency values and Global Reaching Centrality (*GRC*) values for the locally optimal states**. Averaging over the initial full graph (250 initial full graphs and for each initial full graph we have 250 local optimal states) of *N* nodes and the initial subgraphs of $3N$ edges, *p*=*q*=0.2 has been carried out. there is an overall tendency of the PDF-s as a function of the system size. *GRC* and the average efficiency grows with increasing *N*.

In Fig.3 we display a few characteristic dependences of the networks we obtained. Fig. 3(a) shows that larger systems are likely to be more efficient, while Fig. 3(B) shows that for larger networks the optimal configurations seem to fall into two classes with one having a smaller and another one a distinctly more hierarchical structure.

The above results can be qualitatively compared to those obtained for a standard spin-glass model. To do so we have carried out simulations calculating the distribution of the energy of the Edwards-Anderson spin glass model on a square lattice with nearest neighbour interactions[23]. We used a variant with the $J_{ij}$ values being either 1 or -1 with a probability 0.5 and searching for the lowest energy $E = -1/N \sum_{ij} J_{ij} s_i s_j$ by flipping the up or down spins ($s$=1 or $s$= -1) in the nodes. The main point we would like to make here (and display it in Fig. 4) is that in both the E-A and our models the extremal values (minimal energies, maximal efficiencies) corresponding to the various metastable states obtained after optimization are spread, there is a wide selection of locally stable states with varying features.

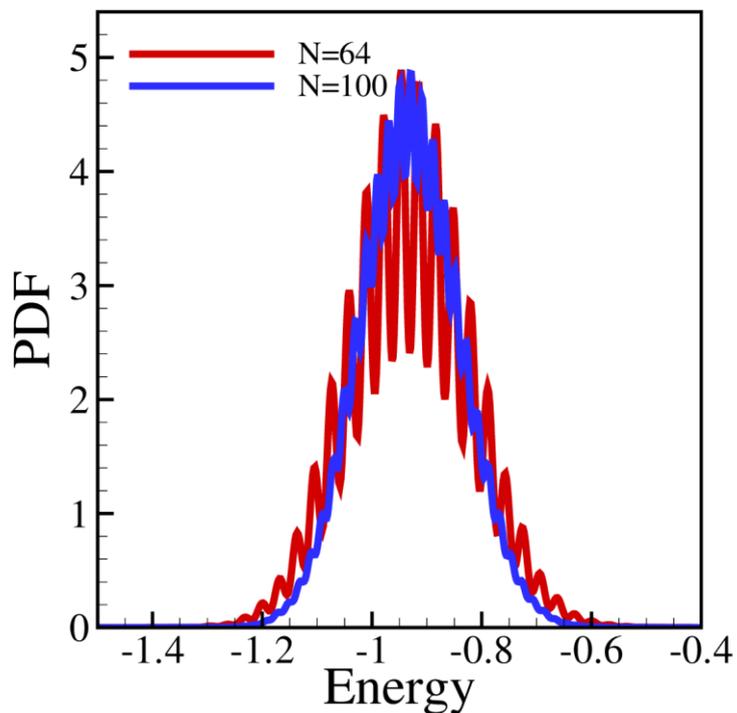

Figure 4. **Distribution of the local minimum of the energy in Edward-Anderson spin glass model for spins being 1 or -1.** Averaging over the 400 different initial conditions and 1000 different local optimal states for each initial condition. The rugged nature of the plots is not due to fluctuations: it follows from the small size of the sample we (intentionally) study.

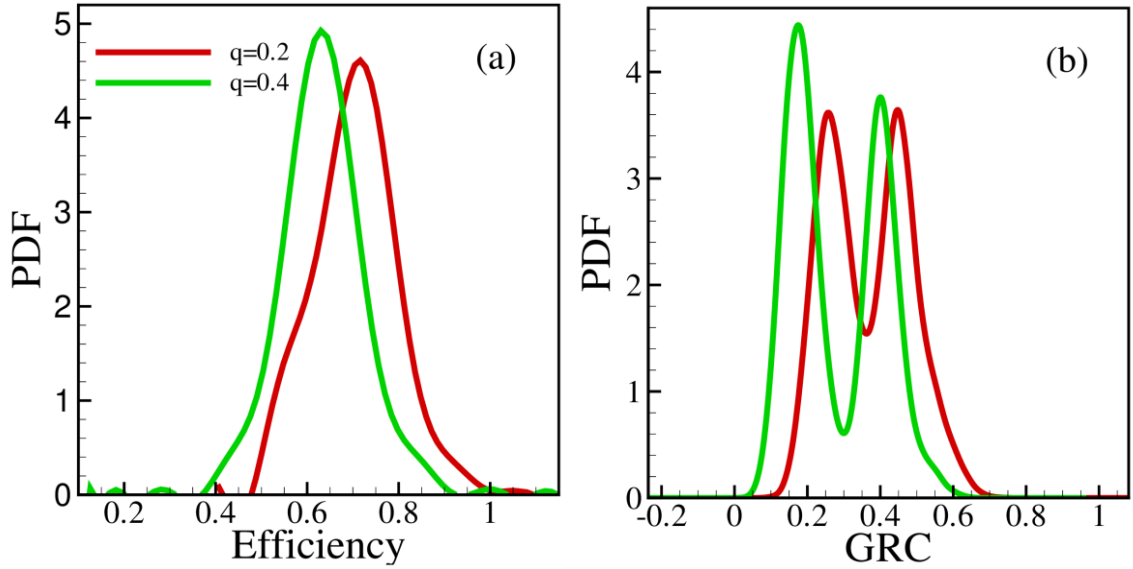

Figure 5. **Comparison of PDF of efficiency and GRC values in local optimal states**. $p$=0.2 while, $q$=0.2 and $q$=0.4 for 64 nodes and $3N$ edges, averaging over initial conditions.

In Fig. 5 we illustrate how the shape of the distributions depend on the parameter $q$ standing for the level of unwillingness to collaborate.

In order to illustrate the variety of optimal structures we obtain, we display in Fig 6. a number of typical examples. These include smaller and larger networks ($N$=16, $N$=128), networks for smaller or larger GRC for $p$=$q$=0.2. For visualization we use the method described in detail in Ref. 23.

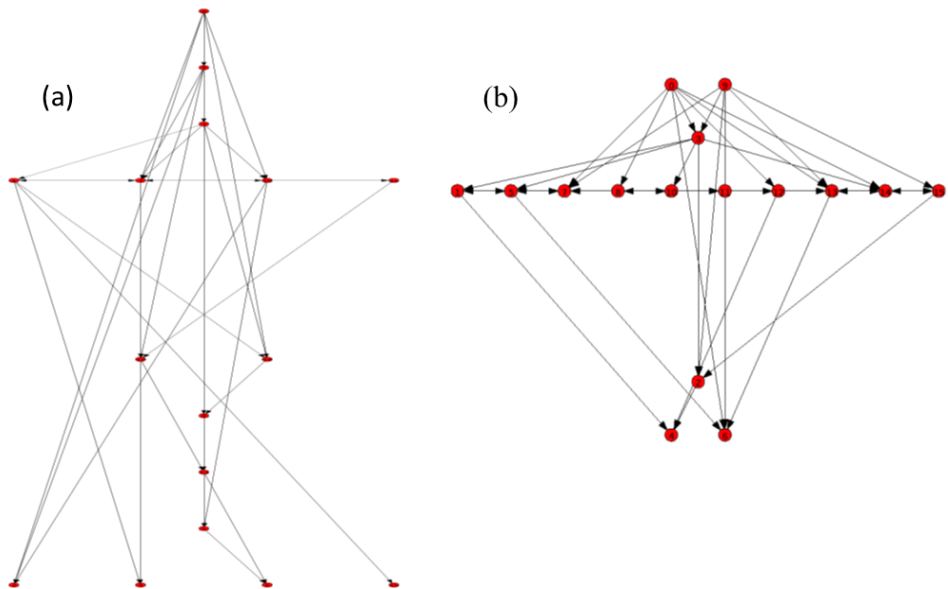

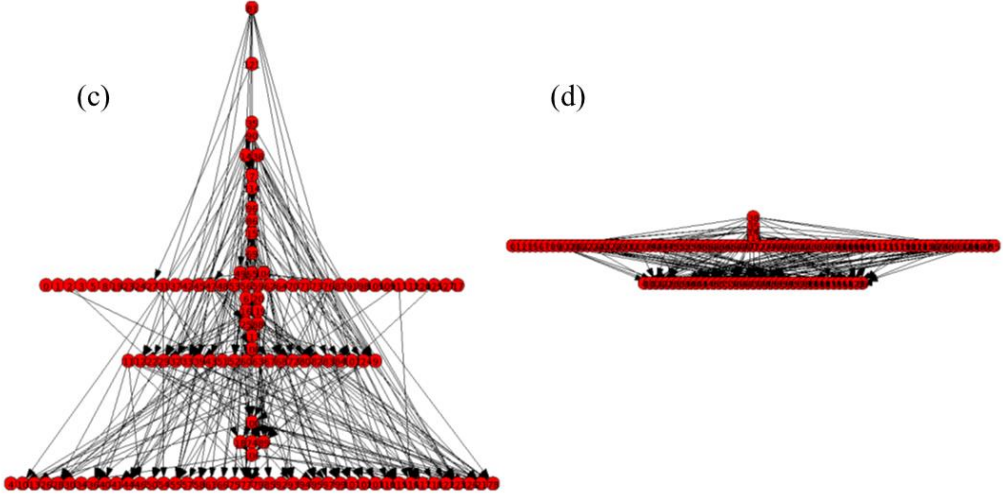

Figure 6. **Hierarchical graphs in selected local optimal states of networks.** $N$=16 and $N$=128, $p$=$q$=0.2, (a) and (c) $GRC$ = 0.62, (b) and (d) $GRC$=0.25

Further interesting features of the emerging networks can be obtained by calculating the dependence of the efficiency on the hierarchy (in general, i.e., the averages of the PDF-s are considered) or the correlations between the number of in-degrees and the out-degrees in the graphs. The results are shown in Fig. 7a. In accord with our expectation, a collaboration network is more efficient if it is more hierarchical. Fig. 7b shows the results we obtained for correlations between the number of incoming and outgoing edges of the nodes (for 250 locally optimal state). These correlations were calculated by using the expression

$$Cor(in, out) = \frac{\langle (k_{in} - \langle k_{in} \rangle)(k_{out} - \langle k_{out} \rangle) \rangle}{[\langle (k_{in} - \langle k_{in} \rangle)^2 \rangle \langle (k_{out} - \langle k_{out} \rangle)^2 \rangle]^{1/2}} \qquad (4)$$

where the summation as well as the averaging (denoted by <...>) is taken over the nodes of the network. According to Fig. 7b the correlations are negative, i.e., nodes with smaller than average incoming (outgoing) edges have in average a larger number of outgoing (incoming) edges. This is also expected for a hierarchical graph of directed edges (with outgoing edges dominating the "top" and incoming edges dominating the nodes at the bottom part of the network.

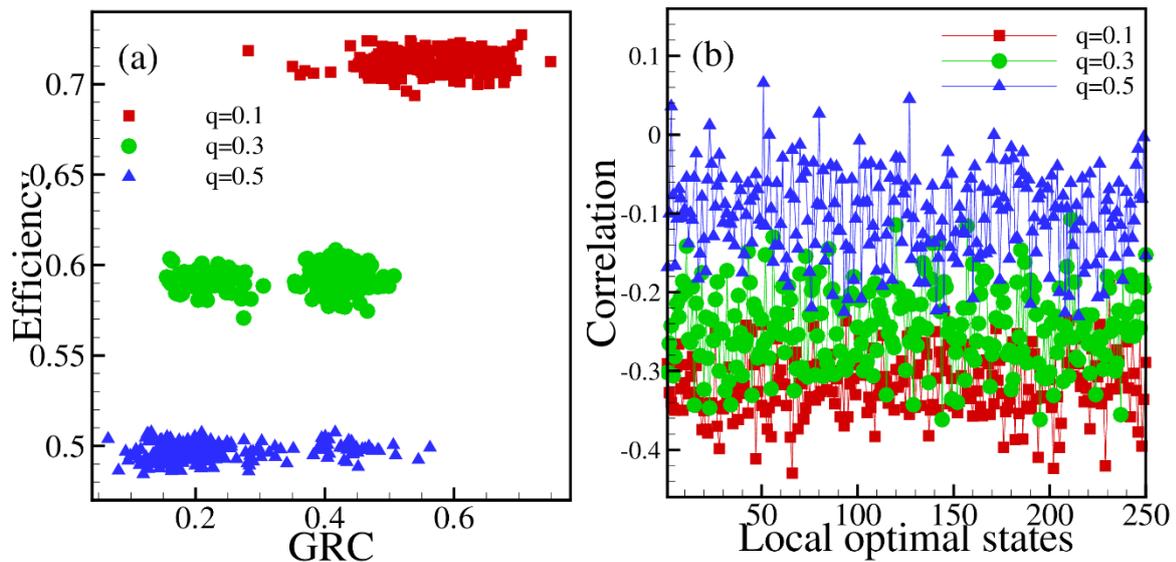

Figure 7. **Features of individual locally optimal states**. (a) Efficiency versus *GRC* values in local optimal states for network with 64 nodes and 3*N* edges starting from a single initial full graph and 250 different subgraphs within this full graph in three different values of *q* and *p*=0.2. (b) Correlation between in-degrees and out-degrees of nodes in local optimal states for three different values of *q* and *p*=0.2 for each parameter set (*N, p, q*) the average efficiency and the level of hierarchy (*GRC* values) can be determined.

Applied to organizations, or, in other words, collaboration networks, our results indicate the following. The structure of these networks is such that they possess the two, perhaps most important features of complex systems: a simultaneous presence of adaptability and stability. Stability is associated with the presence of the local optimum. Only significant perturbations can "kick out" a given arrangement of the participants from this favourable state. However, of the perturbation is large enough (the external conditions change significantly) the network can adapt itself and settle into an alternative, more optimal configuration that suits the new conditions better. The efficiency of the hierarchical structure is higher than a randomly chosen sum of the contributions of the pairwise interactions. These features are in an analogy with those of the glasses including spin glasses. However, our formulation of the system we study is less abstract and takes into account realistic assumptions about the way an organization of collaborating individuals operates. On a wider scale, looking for optimal networks to satisfy given global criteria has a great application potential in many contexts ranging from designing best performing programs to social groups.

**Contributions**

M.Z and T.V. designed the research; M.Z. performed the numerical experiments; T.V. wrote the paper.

**Competing interests**

The authors declare no competing financial interests